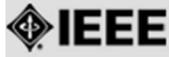

# VOR Adaptation on a Humanoid iCub Robot using a Spiking Cerebellar Model

Note: The following files were submitted by the author for peer review, but cannot be converted to PDF. You must view these files (e.g. movies) online.

Movie S1.mp4
Movie S2.mp4

# VOR Adaptation on a Humanoid iCub Robot using a Spiking Cerebellar Model

Francisco Naveros\*, Niceto R. Luque\*, Eduardo Ros[+], Angelo Arleo[+]

*Abstract* — We embed a spiking cerebellar model within an adaptive real-time control loop able to operate a real robotic body (iCub) when performing different Vestibulo-Ocular Reflex (VOR) tasks. The spiking neural network computation, including event- and time-driven neural dynamics, neural activity and spike-timing dependent plasticity (STDP) mechanisms, leads to a non-deterministic computation time caused by the neural activity volleys encountered during cerebellar simulation. This non-deterministic computation time motivates the integration of a real-time supervisor module able to ensure a well-orchestrated neural computation time and robot operation. Actually, our neurorobotic experimental set-up (VOR) benefits from the biological sensory motor delay between the cerebellum and the body to buffer the computational overloads as well as providing flexibility in adjusting the neural computation time and real-time operation. The real-time supervisor module provides for incremental countermeasures that dynamically slow down or speed-up the cerebellar simulation by either halting the simulation or disabling certain neural computation features (i.e. STDP mechanisms, spike propagation, neural updates) to cope with the real-time constraints imposed by the real robot operation.

This neurorobotic experimental set-up is applied to different horizontal and vertical VOR adaptive tasks that are widely used by the neuroscientific community to address cerebellar functioning. We aim to elucidate the manner in which the combination of the cerebellar neural substrate and the distributed plasticity shapes the cerebellar neural activity to mediate motor adaptation. This work underlies the need for a two-stage learning process to facilitate VOR acquisition.

## I. INTRODUCTION

Descartes' famous "cogito ergo sum" back in the 17th century laid the foundations for creating the dualist body-mind humanistic understanding of human beings. This humanistic conception has resulted in a fruitful dualist tradition in which psychology and neuroscience are two essential partners alternating roles as allies or nemeses. Hence, it is common practice to draw a distinction between the body and the mind when it comes to cognition. Conversely, this mind-centered traditional conception of cognition is now matched by the concept of embodied cognition [1], which emphasizes the coexistence of cognition and body-function as a whole beyond a content-container relationship.

According to this embodied cognition concept, the main aim of the central nervous system (CNS) becomes now to solve and facilitate the interaction of the body with the environment. Consequently, exploring certain capabilities of the nervous system under "behavioral/cognitive tasks" may shed light on how a variety of cellular characteristics, nervous system topologies and/or local synaptic adaptation mechanisms contribute to the body-environment interaction. Nevertheless, the observations of how all these elements play their roles and complement each other are usually not direct or difficult to perform, thus making the hypothesis-experimentation cycle extremely difficult. Computational modeling can partially overcome this limitation by bringing together different sciences such as computational neuroscience, automation or neurorobotics, and by providing easy access to all the elements modeled (neurons, synapses, plasticity mechanisms, etc.).

To that aim, here we have replicated a well-known embodied cognition set-up largely studied in cerebellar neuroscience (the VOR experimental set-up) that can help us to describe and explain what the cerebellum does and does not do using this holistic view. This set-up requires three key elements:

● The vestibulo-ocular reflex (VOR) [2] behavioral experiment set-up (largely used for addressing cerebellar malfunctions) such as our behavioral/cognitive task.

● A cerebellar spiking model such as our neural structure responsible for facilitating the body interaction.

● The humanoid iCub robot [3] such as the front-end human-like body.

Modeling and interconnecting each of these three elements leads us to face state-of-art challenges rarely addressed as a

This work was supported by the "contrato puente" UGR fellowship (F. Naveros), the Juan de la Cierva Spanish fellowship (N. R. Luque), the National Grant (MINECO-FEDER TIN2016-81041-R), by the EU HBP-SGA1, H2020-RIA. 720270 (E. Ros), the French government research program Investissements d'avenir through the Robotex Equipment of Excellence (ANR-10-EQPX-44) and the ANR SilverSight Chair ANR-14-CHIN-0001 (A. Arleo).

F. Naveros, N. R. Luque and E. Ros are with the Research Centre for Information and Communication Technologies (CITIC), Department of Computer Architecture and Technology, University of Granada, Granada, 18014, Spain (e-mail: fnaveros@ugr.es, nluque@ugr.es, eros@ugr.es).

A. Arleo is with Sorbonne Université, INSERM, CNRS, Institut de la Vision, 17 rue Moreau, F-75012, Paris. (e-mail: angelo.arleo@inserm.fr).

\* These two authors contribute equally to the work.
[+] These two authors are Joint Senior Authors.

whole. The next points outline these challenges and how they have been addressed:

(a) Our VOR behavioral protocol, a well-known standardized task among the neuroscientific community, is adopted to draw human-humanoid dis/similarities. Furthermore, it also helps us to frame the dialogue between the cerebellar neural model and the humanoid front-end body in a well-defined experimental set-up.

(b) Our cerebellar model not only operates as the forward controller within the cerebro-cerebellar loop, but it also integrates a variety of neuron models, neural characteristics, and a certain neural system topology that allow us to link bottom-up/top-down observations of the neural cues involved in the body-environment interaction.

(c) Our front-end body mimics certain features of the human body (such as eye movements that can be compensated through a vestibular system during head movements). It is important to remark that the challenge here is to operate this robot using the cerebellar structure in real time (RT). This is a highly demanding task in term of computational efficiency for medium scale neural systems.

Here, we present one of the most comprehensive embodied cognition set-ups. At the core of this experimental set-up, we embed *a RT spiking cerebellar model* in a feed forward control loop able to orchestrate the *vestibulo-ocular adaptation* of the humanoid eye movements mimicking a human being (or standard animal experimental set-up).

## II. MATERIAL AND METHODS

### A. *Behavioral Task: the VOR*

The VOR is a reflexive eye movement that stabilizes the images on the retina during head rotations by producing opposite eye movements that maintain the image in the visual field center (Fig. 1). The VOR depends on the vestibular system, which detects both rotational and translational head movements through the stimulation of semicircular canals and otolithic organs [4].

Rotational VOR (r-VOR) tests compare head and eye velocity movements using VOR gain and phase as markers. The lag between head and eye velocity is called the VOR phase shift (given in degrees), whereas the amplitude ratio of eye and head velocities is called the VOR gain (non-dimensional measurements). For low head rotational frequencies (<5.0 Hz), the VOR gain is close to 1.0 whereas the phase shift is close to 180 degrees [5]. Ideal r-VOR head and eye velocity movements are, therefore, in counter-phase as they are synchronously occurring in opposite directions [2].

Our experimental set-up consists of a 1 Hz r-VOR test in horizontal and vertical planes. Three different sinusoidal amplitudes per plane are evaluated ([30, 90, 150] and [30, 60, 90] deg/s for horizontal and vertical planes respectively).

Cerebellar disorders directly affect the VOR response making VOR tests a powerful tool for cerebellar study and diagnose [6]. The VOR nature is purely feed-forward since it induces prompt compensatory eye movements as a consequence of head movements. The existing mismatches between head movements (signaled by the vestibular organ) and the incoming information to the cerebellum about eye movements represent sensory errors, which are called retinal slips. The forward adaptive control mediated by the cerebellum aims at minimizing these retinal slips.

### B. *Cerebellar Spiking Neural Network Model*

The cerebellar model proposed in this study consists of five neural sub-populations (inspired from [7]): mossy fibers (MFs), granule cells (GCs), inferior olive (IO), Purkinje cells (PCs) and vestibular nuclei (VN) (see Fig. 2). This cerebellar model has been implemented in EDLUT [8-10], an open source event- and time-driven spiking neural simulator mainly oriented to RT embodiment experiments.

The MFs convey the input sensory-motor signals from the eyes and vestibular organ to the cerebellar network. MFs project excitatory afferents onto GCs and VN. The IO cells sense the retinal slip (difference between head and eye velocity movements) and propagate the teaching information through the climbing fibers (CFs) (i.e. IO cell's axons). PCs receive the somatosensory activity from the parallel fibers (PFs) (i.e. GCs' axons) and the teaching signal from the CFs (retinal slip). Finally, the VN close the cerebellar loop with excitatory synapses coming from MFs and IO cells, together with inhibitory synapses coming from PCs. The VN generate the cerebellar output activity, which arrives to the oculomotor neurons (OMNs) responsible for ultimately driving the eye movements.

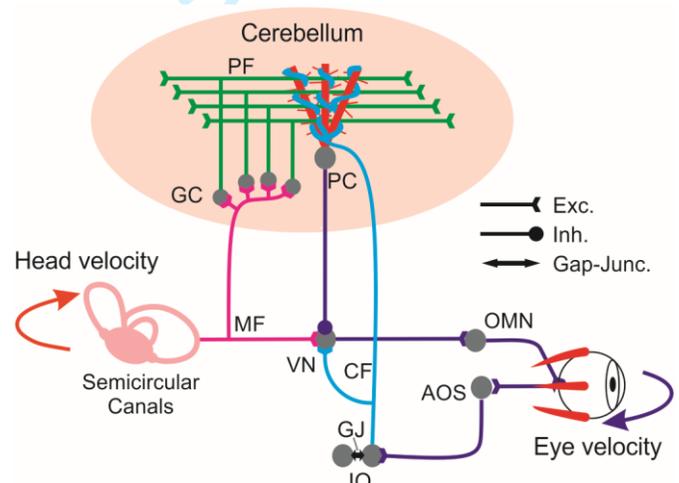

Fig. 2. Vestibular and cerebellar scheme. Connections from semicircular canals in vestibular organ to oculomotor nucleus (OMN) via the flocculus in the cerebellum and the vestibular nuclei (VN), forming the three-neuron reflex arc (MF: mossy fibers, GC: granular cells, PF: parallel fibers, PC: Purkinje cells, CF: climbing fibers, IO: inferior olive, GJ: gap-junction, AOS: accessory optic system).

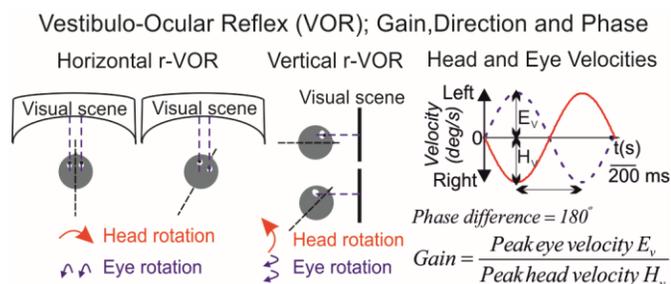

Fig. 1. Vestibulo-Ocular Reflex (VOR) experiment. VOR stabilizes the images on the fovea during horizontal and vertical head rotation tests by producing opposite eye movements that compensate the movement.

- *Mossy fibers (MFs):* 100 MFs are modeled as input neurons able to propagate the sensory-motor information towards GCs and VN. The MF activity is generated by activations of sets of MF neurons following a sinusoidal shape (1 Hz) to encode head velocity movements [11-13], consistently with the functional principles of VOR in cerebellar-control [13]. The total number of activated MFs during an r-VOR trial depends on the head velocity amplitude to be encoded (each MF is sensitive to a small range of velocities).

- *Granular cells (GCs):* 2000 GCs are modeled as a state generator [14-17]. This layer transforms the sensorimotor neural activity coming from the MFs into somatosensory neural activity by generating spatiotemporal patterns that are repeatedly activated during each learning trial (1 s). The passage of time is represented by 500 states that consist of four activated GCs per time-step (2 ms).

- *Purkinje cells (PCs):* 200 PCs are modeled as a single compartment Hodgkin-Huxley (HH) model with five ionic currents (two groups of 100 cells each corresponding to agonist/antagonist muscles). This model is able to replicate the tri-modal spike modes (tonic, silence and bursting [18]) observed in PCs.

- *Inferior olive (IO) cells and climbing fibers (CFs):* 200 IO cells modeled as Leaky Integrate & Fire (LIF) neurons with electrical coupling (two groups of 100 cells each corresponding to agonist/antagonist muscles) conform the olivary system. Each IO cell, through its corresponding CF, makes contact with one PC and one VN cell. Additionally, IO cells are also electrically interconnected via gap-junctions (GJ). The external input activity of IO cells is generated with a probabilistic Poisson process. Given the normalized error signal $\varepsilon(t)$ and a random number $\eta(t)$ between 0 and 1, an IO cell receives an input spike if $\varepsilon(t)>\eta(t)$ [19, 20]. These input stimuli together with the electrical coupling amongst IO cells generate the olivary system activity. Each single CF spike encodes well-timed information regarding the instantaneous error. The probabilistic spike sampling of the error ensures a proper representation of the whole error region over trials, whilst maintaining the CF activity between 1 and 10 Hz per fiber (similar to electrophysiological data [21]). The error evolution can be sampled accurately even at such a low frequency [20, 22].

- *Vestibular nuclei (VN) cells:* 200 VN cells are modeled as LIF neurons (two groups of 100 cells each corresponding for agonist/antagonist muscles). Each VN cell is innervated by an inhibitory afferent from a PC and an excitatory afferent from the CF which simultaneously innervates the same PC. Each VN cell also receives excitatory projections from all MFs (which maintain the baseline VN activity). The spike activity of both VN agonist/antagonist groups is translated into an analog output signal (eye velocity) according to (1-2):

TABLE I
NEURAL NETWORK TOPOLOGY

| Neurons | | | Synapses | | |
|---|---|---|---|---|---|
| Pre-synaptic cells | Post-synaptic cells | Number of synapses | Type | Initial weight (nS) | Weight range (nS) |
| 2000 GC | 200 PC | 400000 | AMPA | 4 | [0, 10] |
| 200 IO | 200 PC | 200 | AMPA | 40 | - |
| 100 MF | 200 VN | 20000 | AMPA | 0 | [0, 1] |
| 200 PC | 200 VN | 200 | GABA | 1.5 | - |
| 200 IO | 200 VN | 200 | AMPA | 1 | - |
| | | | NMDA | 7 | - |
| IO to IO: 5x5 IO neuron squares connected radially from one corner of each 5x5 square to the other three corners | | 320 | GJ | 3 | - |

$$VN_i(t) = \int_t^{t+T_{step}} \delta_{VN_{spike}}(t) \cdot dt \quad (1)$$

$$VN_{output}(t) = \alpha \left( \sum_{i=1}^{N=100} VN_{ag.\ i}(t) - \sum_{j=1}^{N=100} VN_{ant.\ j}(t) \right) \quad (2)$$

where $\alpha$ is the kernel amplitude that normalizes the contribution of each VN cell spike to the cerebellar output correction. This neural topology is summarized in Table I.

C. *Neuron Models*

- *The LIF model (VN)* is implemented according to (3-9). Its neural dynamics is defined by its membrane potential and its excitatory and inhibitory conductances. It is equipped with excitatory (AMPA and NMDA) and inhibitory (GABA) chemical synapses.

$$C_m \frac{dV}{dt} = I_{internal} + I_{external} \quad (3)$$

$$I_{internal} = -g_L \cdot (V + E_L) \quad (4)$$

$$I_{external} = -(g_{AMPA}(t) + g_{NMDA}(t) \cdot g_{NMDA\_INF}) \cdot (V - E_{AMPA}) - g_{GABA}(t) \cdot (V - E_{GABA}) \quad (5)$$

$$g_{AMPA}(t) = g_{AMPA}(t_0) \cdot e^{\frac{(t-t_0)}{\tau_{AMPA}}} \quad (6)$$

$$g_{NMDA}(t) = g_{NMDA}(t_0) \cdot e^{\frac{(t-t_0)}{\tau_{NMDA}}} \quad (7)$$

$$g_{GABA}(t) = g_{GABA}(t_0) \cdot e^{\frac{(t-t_0)}{\tau_{GABA}}} \quad (8)$$

$$g_{NMDA\_INF} = \frac{1}{1 + e^{-62 \cdot V} \cdot \frac{1.2}{3.57}} \quad (9)$$

where $C_m$ denotes the membrane capacitance, $V$ the membrane potential, $I_{internal}$ the internal currents and $I_{external}$ the external currents. $E_L$ is the resting potential and $g_L$ the conductance responsible for the passive decay term towards the resting potential. Conductances $g_{AMPA}$, $g_{NMDA}$ and $g_{GABA}$ integrate all the contributions received by each receptor type (AMPA, NMDA, GABA) through individual synapses. These conductances are defined as decaying exponential functions [8, 23]. Finally, $g_{NMDA\_INF}$ stands for the NMDA activation channel.

- *The LIF model incorporating electrical coupling (IO)* is implemented as the previous LIF model without the NMDA

chemical synapse, but accounting for the electrical synapse as indicated by (10-11).

$$I_{external} = -g_{AMPA}(t) \cdot (V - E_{AMPA}) - g_{GABA}(t) \cdot (V - E_{GABA}) - I_{GJ} \qquad (10)$$

$$I_{GJ} = \sum_{i=N}^{N} w_i \cdot (V - V_i) \cdot \left(0.6 \cdot e^{-\frac{(V-V_i)^2}{50^2}} + 0.4\right) \qquad (11)$$

where $I_{GJ}$ represents the total current injected through the gap-junction (GJ) [24], $w_i$ denotes the synaptic weight between the neuron $i$ and the target neuron, $V$ the target neuron membrane potential, $V_i$ the $i$ neuron membrane potential and $N$ is the total number of input GJ. For a correct operation of the electrical coupling, this model emulates the depolarization and hyperpolarization phases of an action potential by using a threshold process that enables the generation of a triangular voltage function instead of directly resetting the membrane potential each time the LIF neuron fires [25].

- **HH single-compartment model (PC).** This model is based on [26, 27] and consists of a single compartment with five ionic currents and two excitatory (AMPA) and inhibitory (GABA) chemical synapses (12-16).

$$C_m \frac{dV}{dt} = I_{internal} + \frac{I_{external}}{Membrane\ Area} \qquad (12)$$

$$I_{internal} = -g_k \cdot n^4 \cdot (V + 95) - g_{Na} \cdot m_0[V]^3 \cdot h \cdot (V - 50) - g_{Ca} \cdot c^2 \cdot (V - 125) \qquad (13)$$
$$- g_L(V + 70) - g_M \cdot M \cdot (V + 95)$$

$$I_{external} = -g_{AMPA}(t) \cdot (V - E_{AMPA}) - g_{GABA}(t) \cdot (V - E_{GABA}) \qquad (14)$$

$$g_{AMPA}(t) = g_{AMPA}(t_0) \cdot e^{\frac{(t-t_0)}{\tau_{AMPA}}} \qquad (15)$$

$$g_{GABA}(t) = g_{GABA}(t_0) \cdot e^{\frac{(t-t_0)}{\tau_{GABA}}} \qquad (16)$$

where $V$ denotes the membrane potential, $I_{internal}$ the internal currents and $I_{external}$ the external currents. $C_m$ is the membrane capacitance. Conductances $g_{AMPA}$ and $g_{GABA}$ integrate all the contributions received by each chemical receptor type (AMPA and GABA) through individual synapses. These conductances are defined as decaying exponential functions [8, 23]. Finally, $g_K$ is a delayed rectifier potassium current, $g_{Na}$ a transient inactivating sodium current, $g_{Ca}$ a high-threshold non-inactivating calcium current, $g_L$ a leak current, and $g_M$ a muscarinic receptor suppressed potassium current.

The dynamics evolution of each gating variable ($n$, $h$, $c$, and $M$) can be computed using the following differential equation:

$$\dot{x} = \frac{x_0[V] - x}{\tau_x[V]} \qquad (17)$$

where $x$ indicates the variables $n$, $h$, $c$, and $M$. The implemented equilibrium function is determined by the term $x_0[V]$ and time constant $\tau_x[V]$ (Table II).

The sodium activation variable has been replaced and approximated by its equilibrium function $m_0[V]$. The M-current presents a temporal evolution significantly slower than the rest of variables that allows the PC trimodal spike modes named burst, silence and tonic. For the sake of computational efficiency, $I_K$ and $I_{Na}$ currents can be substituted by a simple threshold process that triggers the generation of a triangular voltage function each time the neuron fires [25]. This triangular voltage depolarization drives the state of ion channels similarly to the original voltage depolarization during the spike generation. The final internal current is:

$$I_{internal} = -g_{Ca} \cdot c^2 \cdot (V - 125) - g_L(V + 70) - g_M \cdot M \cdot (V + 95) \qquad (18)$$

TABLE II
IONIC CONDUCTANCE KINETIC PARAMETERS

| Cond. type | Steady–state Activation/Inactivation | Time constant (ms) |
|---|---|---|
| $g_K$ | $x_0[V] = \dfrac{1}{1 + e^{\frac{-V-29.5}{10}}}$ | $\tau_x[V] = \begin{cases} 0.25 + 4.35 e^{\frac{V+10}{10}}; \text{ if } V \leq 10 \\ 0.25 + 4.35 e^{\frac{-V-10}{10}}; \text{ if } V > 10 \end{cases}$ |
| $g_{Na}$ | $x_0[V] = \dfrac{1}{1 + e^{\frac{V-59.4}{10.7}}}$ | $\tau_x[V] = 0.15 + \dfrac{1.15}{1 + e^{\frac{33.5}{15}}}$ |
| $m_0[V]$ | $m_0[V] = \dfrac{1}{1 + e^{\frac{-V-48}{10}}} m$ | |
|  | **Forward Rate Function ($\alpha$)** | **Backward Rate Function ($\beta$)** |
| $g_{Ca}$ | $\alpha = \dfrac{1.6}{1 + e^{-0.0072(V-5)}}$ | $\beta = \dfrac{0.02(V + 8.9)}{e^{\frac{V+8.9}{5}} - 1}$ |
| $g_M$ | $\alpha = \dfrac{0.3}{1 + e^{\frac{(-V-2)}{5}}}$ | $\beta = 0.001 e^{\frac{(-V-70)}{18}}$ |
|  | **Steady–state Activation/Inactivation** | **Time constant(ms)** |
|  | $x_0[V] = \dfrac{\alpha}{\alpha + \beta}$ | $\tau_x[V] = \dfrac{1}{\alpha + \beta}$ |

D. *Synaptic Plasticity*

The overall input-output function of the cerebellar network model is made adaptive through STDP mechanisms at different sites. These STDP mechanisms balance long-term potentiation (LTP) and long-term depression (LTD) (see [7] for an in-depth review of the implemented synaptic mechanisms).

- **PF–PC synaptic plasticity:** The LTD/LTP balance at PF–PC synapses is based on (19-20):

$$LTD\ \Delta w_{PF_j - PC_i}(t) = \alpha \cdot \int_{-\infty}^{CF_{spike}} k\left(\frac{t - t_{CF_{spike}}}{\tau_{LTD}}\right) \cdot \delta_{PF_{spike}}(t) \cdot dt \qquad (19)$$

$$LTP\ \Delta w_{PF_j - PC_i}(t) = \beta \cdot \delta_{PF_{spike}}(t) \qquad (20)$$

where $\Delta W_{PFj-PCi}(t)$ denotes the weight change between the $j^{th}$ PF and the target $i^{th}$ PC; $\tau_{LTD}$ = 100 ms is the time constant that compensates the sensorimotor delay; $\delta_{PF}$ is the Dirac delta function corresponding to an afferent spike from a PF; $\alpha$ = -0.0304 nS is the synaptic efficacy decrement; $\beta$ = 0.0184 nS is the synaptic efficacy increment; and the kernel function $k(x)$ [7] is defined as:

$$k(x) = e^{-x} \cdot \sin(x)^{10} \qquad (21)$$

The STDP rule [28] defined by (19) produces a synaptic efficacy decrement (LTD) when a spike from the CF reaches the target PC neuron. The amount of synaptic decrement depends on the activity arrived through the PFs. This activity is convolved with the integrative kernel defined in (21) and multiplied by the synaptic decrement $\alpha$. The effect on the presynaptic spikes arriving through PFs is maximal over the 100 ms time window before the postsynaptic CF spike arrival, thus accounting for the sensorimotor pathway delay [20, 29-31]. The amount of LTP at PF–PC synapses is fixed (20), with an increase in synaptic efficacy equals to $\beta$ each time a spike arrives through a PF to the targeted PC.

- **MF–VN synaptic plasticity:** The LTD/LTP dynamics at MF – VN synapses is based on (22-23):

$$LTD\ \Delta w_{MF_j-VN_i}(t) = \alpha \cdot \int_{-\infty}^{\infty} k\left(\frac{t - t_{PC_{spike}}}{\sigma_{MF-VN}}\right) \cdot \delta_{MF_{spike}}(t) \cdot dt \quad (22)$$

$$LTP\ \Delta w_{MF_j-VN_i}(t) = \beta \cdot \delta_{MF_{spike}}(t) \quad (23)$$

with $\Delta W_{MFj-VNi(t)}$ denoting the weight change between the $j^{th}$ MF and the target $i^{th}$ VN; $\sigma_{MF-VN} = 5$ ms standing for the time width of the kernel; $\delta_{MF}$ representing the Dirac delta function that defines a MF spike; $\alpha = -0.002048$ nS is the synaptic efficacy decrement; $\beta = 0.000792$ nS is the synaptic efficacy increment; and the integrative kernel function $k(x)$ [7] defined as:

$$k(x) = e^{-|x|} \cdot \cos(x)^2 \quad (24)$$

The STDP rule defined by (22) produces a synaptic efficacy decrease (LTD) when a spike from the PC reaches the targeted VN neuron. The amount of synaptic decrement depends on the activity arrived through the MFs. This activity is convolved with the integrative kernel defined in (24) and multiplied by the synaptic decrement $\alpha$. This LTD mechanism considers those presynaptic/postsynaptic MF spikes that arrive before/after the postsynaptic/presynaptic PC spike arrival within the time window defined by the kernel ($\sigma_{MF-VN}$). The amount of LTP at MF - VN synapses is fixed, with an increase in synaptic efficacy equals to $\beta$ each time a spike arrives through a MF to the targeted VN.

E. *VOR Plant. Modeling the Mechanical Circuitry*

The cerebellum operates as a biological feed-forward controller within a control loop. The cerebellar output is meant to drive adaptation from the vestibular nuclei through a set of motor neurons, nerve fibers and muscles finally to the eye. This VOR mechanical pathway is modeled (within EDLUT) as a VOR mechanical circuitry which is identified as a continuous-time mathematical model with two poles:

$e(kT), E(s): eye\ motion\ (output)$ (25)

$h(kT), H(s): head\ motion\ (input)$ (26)

$$VOR(s) = \frac{E(s)}{H(s)} = \frac{K \cdot T_{C1} \cdot s}{(T_{C1} \cdot s + 1) \cdot (T_{C2} \cdot s + 1)} \cdot e^{-s\tau_{delay}} \quad (27)$$

There are four parameters in the model: $Q=[K, T_{C1}, T_{C2}, \tau_{delay}]$. The delay parameter $\tau_{delay}$ captures the delay that exists in communicating the signals from the inner ear to the brain and eyes. Based on the number of synapses involved in the VOR, this delay is estimated to be around 5 ms [32, 33]. The gain parameter $K$ models the fact that the eyes do not perfectly cope with the head movement. This parameter is assumed to be between 0.6 and 1 [32, 33]. The $T_{C1}$ parameter represents the dynamics associated with the semicircular canals as well as some additional neural processing. The canals are high-pass filters, because after a subject has been put into rotational motion, the neural active membranes in the canals slowly relax back to resting position, so the canals stop sensing motion. Based on the mechanical characteristics of the canals, combined with additional neural processing which prolongs this time constant to improve the accuracy of the VOR, the $T_{C1}$ parameter is assumed to be between 10 and 30 seconds [32, 33]. Finally, the $T_{C2}$ parameter captures the oculomotor plant dynamics, i.e. the eye, muscles and tissues attached to it. The $T_{C2}$ parameter is assumed to be between 0.005 and 0.05 seconds.

The temporal response for the VOR transfer function requires calculating the inverse Laplace transform, thus obtaining (28-29) (note that the delay is modeled and inserted within the control loop).

$$\begin{bmatrix} \dot{x}_1 \\ \dot{x}_2 \end{bmatrix} = \begin{bmatrix} 0 & 1 \\ -a_0 & -a_1 \end{bmatrix} \cdot \begin{bmatrix} x_1 \\ x_2 \end{bmatrix} + \begin{bmatrix} 0 \\ h(t) \end{bmatrix} \quad (28)$$

$$y = [b_0\ \ b_1] \cdot \begin{bmatrix} x_1 \\ x_2 \end{bmatrix} \quad (29)$$

Where:

$$a_0 = \frac{1}{T_{C1} \cdot T_{C2}};\ a_1 = \frac{(T_{C1} \cdot T_{C2})}{T_{C1} \cdot T_{C2}};\ b_0 = 0;\ b_1 = \frac{K \cdot T_{C1}}{T_{C1} \cdot T_{C2}} \quad (28)$$

The VOR plant model parameters are adjusted using a genetic algorithm to fit experimental and clinical observations [32-34]. The resulting parameters values are $k=1.0$, $T_{C1}=15$, $T_{C2}=0.05$.

F. *iCub Robot*

The humanoid iCub robot, used as front-end body, can sense its own body position (proprioception) and movement (using accelerometers and gyroscopes) [3]. We control the simulated (in gazebo [35]) and actual iCub robot using YARP: Yet Another Robot Platform [36]. This API enables the sending of motor commands and the receiving of sensory information from the robot.

The r-VOR protocol only requires moving the iCub head (controlled by the neck) and eyes. Both the neck and eyes consist of a serial chain of rotations with 3 degrees of freedom. Additionally, the eyes incorporate a camera.

Prior to the real iCub robot implementation, we have validated and calibrated the r-VOR protocol using the virtual iCub version.

G. *Control Loop*

The sensory-motor information needs to flow between the humanoid robot and the cerebellar neural network causing effects on one another. Figure 3 shows how this interaction is managed by two key elements; the inner/outer control loop (including the cerebellar model and the VOR plant implemented in EDLUT) and the robot interface (using YARP to connect with the robot). Both elements have been

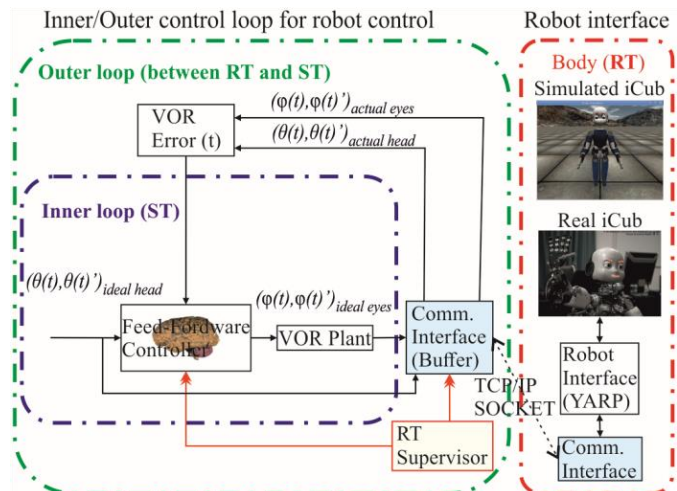

Fig. 3. Cerebellar inner/outer control loop and robot interface for r-VOR tasks. The vestibular and teaching (error) neural signals arrive to the cerebellar model through the MFs and IOs respectively (after an analog-to-spike conversion). The cerebellar model processes this input activity and generates the corresponding output response through the VN. This VN output spike activity is transformed into an analog signal that feeds the VOR plant. The VOR plant output, representing the eye velocity commands, is stored in the outer loop buffer. The robot interface module concurrently connects with the outer loop communication interface, receiving the eye velocity commands from the inner loop and sending the sensory information from the iCub sensors (head and eyes trajectories). Then, the outer loop compares both trajectories, generating the teaching signal (IO activity) in the VOR error module. The RT supervisor manages the simulation speed and the synchronization between both elements using a temporal buffer.

independently designed and interconnected via TCP/IP. This modular structure presents two main advantages:

a) The control loop may operate any humanoid robot as long as the robot interface program is adjusted.

b) The computational load of the control loop operation may be decoupled from the robot interface program (both running in different CPU cores/computers when needed).

The inner/outer control loop configuration also permits to uncouple the execution time of the iCub internal clock, operating in RT at the robot interface, from the execution time of the cerebellar internal clock operating in simulation time (ST), at the inner loop. The simulation time speed (ratio between execution and simulation time) at the robot interface is always 1 for a real robot (RT restriction imposed by the robot). In contrast, the cerebellar simulation time speed at the inner loop may be less than, greater than or equal to 1 per each simulation time step, depending on the neural activity to be processed. This control configuration uses the outer loop as a temporal buffer between the iCub internal clock at the robot interface and the cerebellar internal clock at the inner loop. Since both elements run in different time domains, a RT supervisor operating as a man-in-the-middle is mandatory. The RT supervisor controls the inner loop simulation speed (ratio between ST and RT) as well as the communication period between the cerebellar control loop and the iCub robot to ensure a coherent sensor-actuator time accessing.

The temporal difference between the simulation time of the iCub internal clock and the cerebellar internal clock offers the opportunity to pre-compute and store the neural activity of the cerebellar model prior to RT without neural information losses

(the cerebellar simulation time $ST_{net}$ may be equal to or greater than the iCub simulation time $ST_{rob}$). This difference between simulation times is limited by the sensory-motor pathway delay (upper bound). This sensory-motor delay (100 ms) includes the time period elapsed from the sensory information reception to, information transmission along nerve fibers, neural processing time responses and the final motor output response [37]. Since our technological delay (15 ms) is significantly shorter than the sensory-motor pathway delay, there is a total of 85 ms temporal difference for neural pre-computation (the cerebellar simulation time $ST_{net}$ must not exceed in more than 85 ms the iCub simulation time $ST_{rob}$). This behavior is summarized in (29-31).

$$ST_{rob} \leq ST_{net} \leq ST_{rob} + 85ms \qquad (29)$$

$$ST_{rob} = RT \qquad (30)$$

$$RT \leq ST_{net} \leq RT + 85ms \qquad (31)$$

The RT supervisor manages this temporal difference to ensure that the neural computational load fluctuations encountered during simulation due to volleys of neural activity can be coped with a RT execution.

H. *Real-Time Supervisor*

The EDLUT simulator incorporates a hybrid event- and time-driven simulation scheme [8, 9, 38]. EDLUT takes full advantage of parallel processing (in CPU and GPU) for neural layers with high levels of neural activity adopting a time-driven simulation scheme (i.e. the PC layer). EDLUT also takes full advantage of parallel processing (in CPU) for neural layers with sparse neural activity adopting an event-driven simulation scheme (i.e. the GC layer). This hybrid simulation scheme significantly optimizes the simulation speed although it remains inherently non-deterministic due to the volleys of neural activity encountered. This non-deterministic behavior demands the development and integration of a RT supervisor.

The RT supervisor, developed here for EDLUT, ensures that the simulation time of the neural activity ($ST_{net}$) copes with the iCub RT internal clock and does not surpass the temporal difference between the biological and artificial pathway delays of 85 ms (31). The RT supervisor deploys a set of gradual countermeasures depending on the time distance between $ST_{net}$ and RT, thus affecting the neural computation to a lesser or greater degree (Table III). The smaller the time distance between $ST_{net}$ and RT, the higher the contingency level and the more drastic the countermeasures are.

During r-VOR tests, the RT countermeasures taken are usually minimal (0-1 contingency levels). However, under occasional large neural computational loads, the countermeasures taken may range from 2 to 4. Under these conditions, EDLUT disengages some neural computation elements, thus causing a slight degradation over the final cerebellar outcome. A permanent contingency level value between 2 and 4 means that EDLUT does not meet the neural dynamic computation requirements. The outcome, therefore, would drastically differ from what was expected. We monitor

TABLE III
REAL-TIME COUNTERMEASURES

| Cont. level | Contingency tasks |
|---|---|
| 0 | The time distance between $ST_{net}$ and RT is too close to 85 ms). The neural simulation has to be halted. |
| 1 | Standard simulation. No countermeasures are needed. |
| 2 | The time distance between $ST_{net}$ and RT is near zero. Learning rules are disengaged to speed-up the neural simulation. |
| 3 | The time distance between $ST_{net}$ and RT is even closer to zero. Spikes propagation and neuron model updates are also disengaged to further speed-up the neural simulation. |
| 4 | The time distance between $ST_{net}$ and RT is too close to zero. All the non-vital neural dynamic computation is disengaged (i.e. internal spike generation, periodic weight saving operation, etc.). |

these countermeasure levels during all the simulations to validate the obtained results.

## III. RESULTS

The behavioral tasks proposed as a test-bed consists of a robotic r-VOR cerebellar adaptation process performed in the horizontal and the vertical planes. The r-VOR cerebellar adaptation process lasts for 300 trials, one second per trial in RT.

### A. *The Cerebellar Adaptation Process*

The STDP mechanisms located at PF-PC and MF-VN afferents (see Methods) modulate the cerebellar output response during r-VOR adaptation. Figure 4 depicts how the cerebellar neural activity evolves during synaptic adaptation (Fig. 5). The cerebellar input signals from the vestibular organ (Fig. 4, left column) remains unchanged, as the horizontal head rotation movement during the adaptation process. This sensory input activity propagated by the MFs is transformed into a sparse neural coding at GCs, which represents univocally the passage of the time by a set of spatiotemporal neural patterns repeatedly activated during each learning trial (Fig. 4, left column). The PF-PC STDP mechanism correlates the GC neural activity (propagated by the PFs) with the teaching signal (error signal) sensed by the CFs by modifying the synaptic weights at this site (Figs. 5A and B). Once the VOR adaptation is accomplished at PF-PC synapses, it is then transferred (in counter phase due to the inhibitory nature of PC axons) and consolidated in deepest cerebellar structures (MF-VN) (Figs. 5C and D), consistently with the *two learning stages hypothesis* proposed by [39] (see [7] for an in-depth review).

At the beginning of the learning process, the cerebellum starts with a blank sheet and adapts PF-PC (4nS) and MF-VN (0nS) synaptic weights from scratch (Figs. 5A and C). CF activations are maximal (10 Hz corresponding to the maximal error sensed) (Fig. 4A, central column) and the cerebellar output is negligible since the adaptation process is not yet deployed (Fig. 4A, right column); the eyes are moving conjointly with the head.

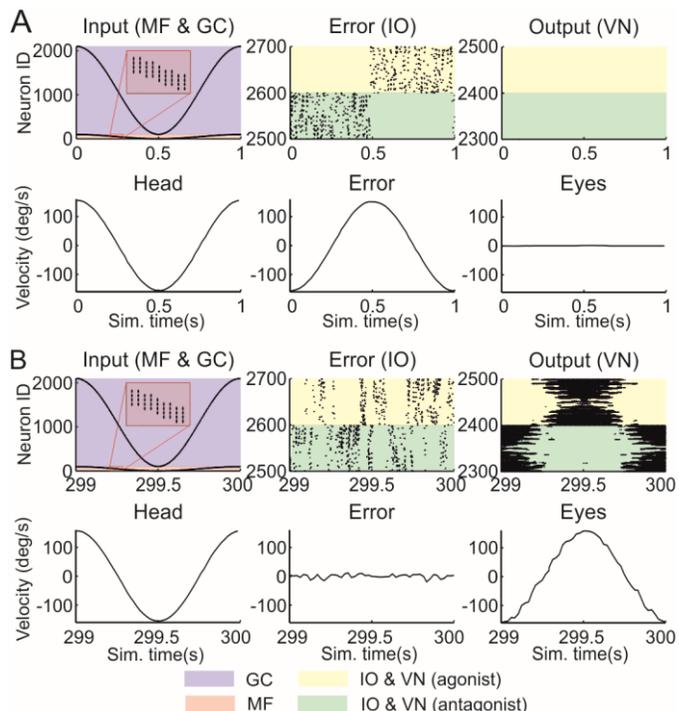

Fig. 4. Cerebellar input/output signals during horizontal r-VOR task (150 deg/s). A) Initial learning stage. B) Final learning stage. The first row of each panel depicts the spiking cerebellar input/output signals, whereas the second row shows their analog translation. The left column depicts the cerebellar input (head velocity), the central column the error input (mismatch between head and eyes velocities) and the right column the cerebellar output (eyes velocity with respect to head movement). The right column analog signal actually represents the cerebellar output emerging from the VOR plant.

At the end of the learning process, the cerebellar output (Fig. 4B, right column) fully compensates for the head movement thanks to the adaptation process deployed (Figs. 5B and D). The resulting CF activations are minimal (1-2 Hz corresponding to the CF baseline activation in the absence of error sensed) (Fig 4B, central column). The zenith view of the synaptic weight distribution at PF-PC and MF-VN synapses depict the footprints that the STDPs generate; footprints that are the photo negative of one another (Figs. 5B and D). Figures 5B and D also show two differentiated areas representing the final agonist/antagonist microzone balance.

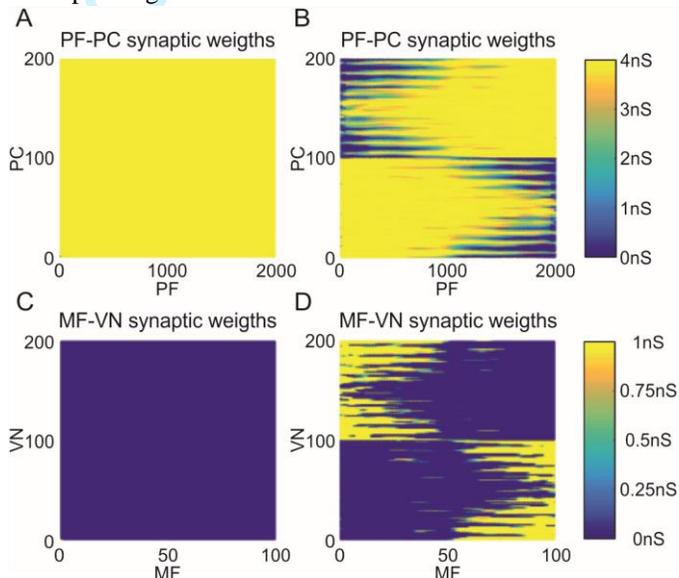

Fig. 5. Synaptic weight distribution during a horizontal r-VOR task (150 deg/s). The first row depicts synaptic weight distribution at the PF-PC innervations. The second row depicts the synaptic weight distribution at the MF-VN innervations. For each row, the left hand column represents the synaptic weight distoutions at the initial learning stage, whilst the right hand column represents the synaptic weight distributions at the final learning stage.

The combination of PF-PC afferents, whose synaptic weights are saturated up (Fig. 5B), and the matching MF-VN afferents saturated down (Fig. 5D), cause a maximal PC inhibitory action over VN cells accompanied by a reduction of the VN cell activity through the MFs (Fig. 4B, right hand column). The combination of these afferents minimally contributes to cerebellar adaptation. In contrast, the combination of PF-PC afferents whose synaptic weights are saturated down (Fig. 5B), and the matching MF-VN afferents saturated up (Fig. 5D), causes a minimal PC inhibitory action accompanied by an augmentation on the vestibular activity through MFs. The combination of these afferents greatly contributes to the cerebellar adaptation. The multiple PF-PC/MF-VN synaptic weight combinations between these two extreme cases provide for a ranged contribution of a set of afferents to cerebellar adaptation.

We expand the r-VOR cerebellar adaptation capabilities to three different amplitudes per plane (horizontal and vertical plane). Figure 6 shows in the left hand column the mean absolute error (MAE) evolution between movements of the head and eyes. This measure helps us to evaluate r-VOR accuracy during the 300 trials needed for fully deploying cerebellar learning. MAE is always maximal at the beginning of the learning process (during the first trials), when no cerebellar adaptation is deployed yet. The MAE progressively decreases as the cerebellar adaptation takes over. Once the synaptic weight distributions at PFs-PCs and MFs-VN are settled and stabilized, the MAE converges to its minimal; the larger the head velocity amplitude to compensate, the greater the cerebellar compensatory output and the time to obtain the optimal synaptic weight distribution are (the MAE requires more time to converge).

Figure 6 also shows in the right hand column the velocity curves obtained for the head and eyes during the last trial of each r-VOR task. Both the head and eyes curves are similar in amplitude (VOR gain close to 1) but in counter-phase (VOR phase close to 180º). These results are consistent with empirical observations [5].

### B. *Meeting RT Requirements, RT Supervisor Process*

The RT supervisor implemented within the inner/outer control loop is able to control the simulation time speed of the cerebellar neural activity to meet the RT bounds (see methods), thus enabling the sensory-motor information to flow in both ways (cerebellar model to iCub robot and vice versa). Figure 7A depicts the temporal evolution of the cerebellar simulation time speed (ratio between execution and simulation time) corresponding to two horizontal r-VOR tasks (150 deg/s) with and without the RT supervisor engaged. This comparison has been performed using the simulated iCub robot.

The cerebellar simulation time ($ST_{net}$) without the RT supervisor engaged (no RT version) overpasses the RT bounds of (31), making a coherent sensory-motor information propagation in the real robot impossible (Figs. 7A and B). In contrast, the cerebellar simulation time with the RT supervisor engaged (RT version) is slowed down or speeded-up to cope with the RT bounds of (31), making possible a coherent sensory-motor information propagation in the real robot (Figs. 7A and C). The RT supervisor, using the RT contingency levels described in Table III, ensures that the time distance between the simulation time of the cerebellar neural activity ($ST_{net}$) and RT is always between the bounds required for the robot communication. When this time distance is close to the upper RT bound (Fig. 7C), the contingency level switch to 0 (Fig. 7D), halting the cerebellar simulation. On the contrary, when the distance is close to the lower RT bound (Fig. 7C), the contingency level takes values of between 2 and 4 (Fig. 7D), progressively disengaging different neural elements to speed up the cerebellar simulation.

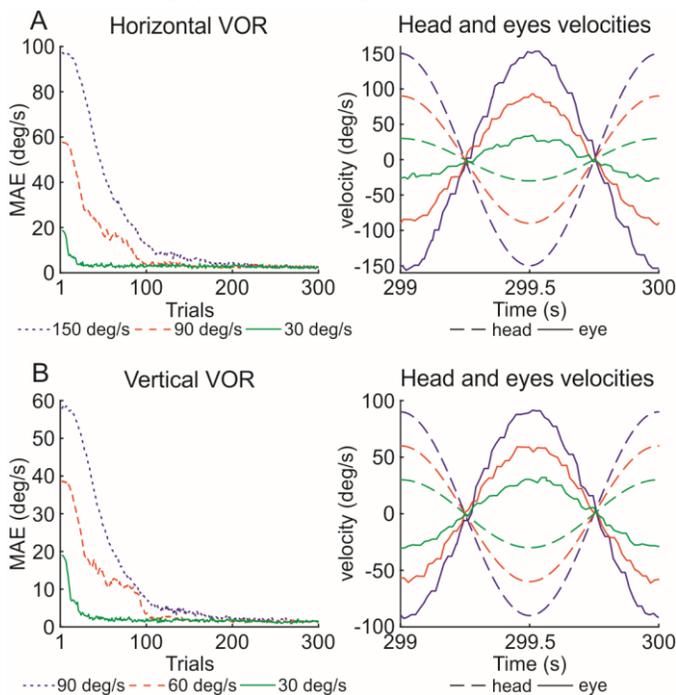

Fig. 6. Cerebellar output response to horizontal and vertical r-VOR tasks. A) Horizontal r-VOR tasks (150, 90 and 30 deg/s). B) Vertical r-VOR tasks (90, 60 and 30 deg/s). The left hand column represents the mean absolute error (MAE) evolution between the velocities of the head and eyes during the 300 trials (1 second per trial) of each experiment. The right hand column represents curves of the head and eyes during the last trial of each task.

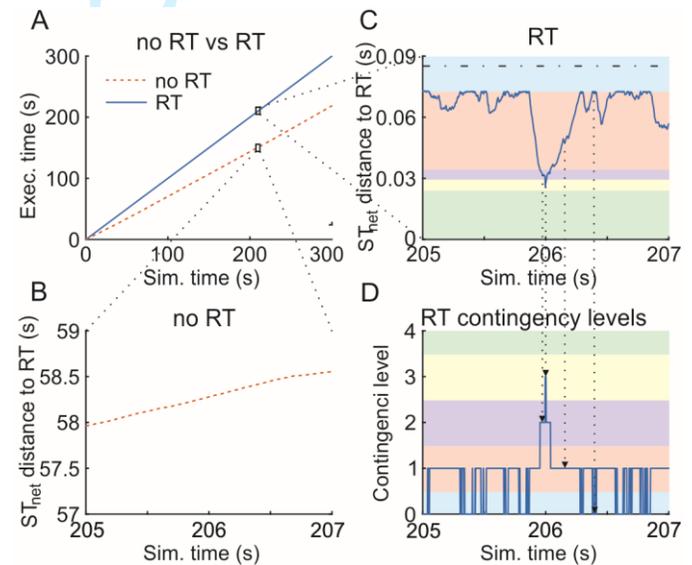

Fig. 7. RT supervisor impact on the neural simulation time speed in a horizontal r-VOR task (comparison when enabling and disabling the RT supervisor). A) Cerebellar simulation time speed (ratio between simulation and execution time). B) and C) Cerebellar simulation time ($ST_{net}$) distance to RT. This distance must not surpass 85 ms to meet the RT bounds required for a coherent sensory-motor propagation. D) RT supervisor contingency level depending on the distance between $ST_{net}$ and RT.

Table IV shows the time spent at each contingency level in our RT simulation. The time spent at contingency levels 0-1 is 98.69% of the total simulation time (standard simulation without degradation). The time spent at contingency level 2 is 1.2693%. Level 2 involves a slight degradation in the neural computation caused by the disengagement of the STDP mechanisms (the learning process is delayed). Finally, the time spent at contingency levels 3-4 is 0.0407%. Levels 3-4 involve a larger degradation in the neural computation due to the disengagement of critical neural elements, i.e. spike generation and propagation.

To measure the neural degradation impact on the RT simulation, we calculate the mean and standard deviation of the difference between the synaptic weight distributions at PF-PC and MF-VN obtained for the RT and no RT simulation. These values are -0.0059±0.0523 nS for PF-PC synapses and 0.00003±0.0776 nS for MF-VN synapses. We also calculate the mean and standard deviation of the difference between MAE evolutions of the cerebellar output response for the RT and no RT simulation, thus obtaining 0.1165±0.3293 deg/s. These deviations are negligible, which makes the RT supervisor impact in the neural outcome minimal.

### C. *Robotic r-VOR cerebellar adaptation. Proof of concept*

Two movies are included to visually verify the entire adaptation of the reflex as supplementary material. Movie S1 shows the evolution of cerebellar adaptation process in a simulated iCub robot whereas movie S2 shows the real iCub robot. Each movie includes the six r-VOR tasks proposed (three in the horizontal plane and three in the vertical plane) and compares the cerebellar initial learning stages with the final stages. Figure 8 shows a snapshot of both movies for the

TABLE IV
REAL-TIME SUPERVISOR IMPACT

| Cont. level | Percentage (%) | Degradation |
|---|---|---|
| 0 | 28.9847 | Halted simulation without degradation |
| 1 | 69.7053 | Standard simulation without degradation |
| 2 | 1.2693 | Degraded simulation delaying leaning process |
| 3 | 0.0267 | Degraded simulation removing neural propagation |
| 4 | 0.0140 | Degraded simulation removing neural generation and propagation |

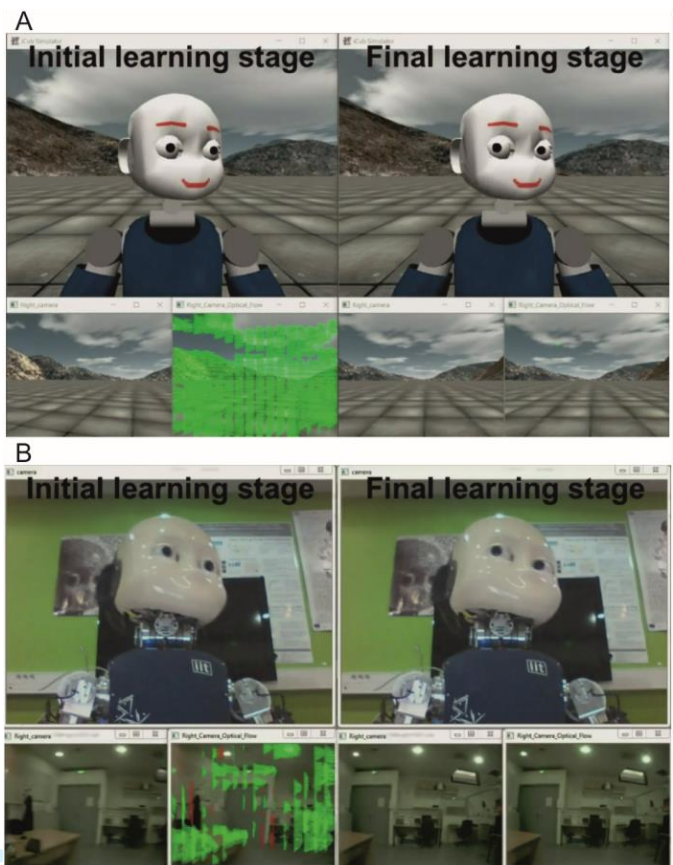

Fig. 8. Snapshot of the movies filming the r-VOR in the simulated (A) and real (B) iCub robot. The left hand windows represent the initial learning stage whereas the right hand windows represent the final learning stage. The upper windows in A and B show the head and eyes movements whereas the lower windows show the images filmed by the eye cameras. The optical flow is computed over the camera images (superimposed in green arrows) indicating quantitatively the level of stabilization of the filmed image on the "retina".

150 deg/s horizontal r-VOR task. Each movie consists of six windows. The windows on the left show the initial learning stage whilst the windows on the right show the final learning stage. The windows at the top show the iCub robot moving the head (reference signal) and eyes (controlled signal). The windows at the bottom show the images captured by the eye cameras that are stabilized at the end of the learning process. We plot the optical flow over the images captured by the eye cameras (green and red arrows). These arrows give us an estimate of the relative motion of the visual scene observed by the iCub robot (i.e. they are retinal slip proxies). At the beginning of the learning process, the eyes and head move conjointly and the optical flow is maximal since the image is not stable in the "retina". At the end of the learning process, the cerebellar model is able to compensate for the head rotation, the eye camera images are stabilized and the optical flow is significantly reduced.

## IV. DISCUSSION

Different artificial intelligence VOR solutions in robotic platforms have been proposed during the last decade to try to give a better insight into the computational primitives underneath our CNS. These solutions are organized into two families according to their biological plausibility: machine learning and the cerebellar-based family.

A. *The machine learning family*

The embodied cognition approach (VOR) is solved without devoting attention to the biological restrictions imposed by the neural structures within the nervous system. The algorithms mediating the cerebellar role operation are claimed to be inspired in either the cerebellar architecture or the cerebellar functionality or both. However, the parallelisms to be drawn between the cerebellar operation/architecture and the algorithms proposed are generally constrained to a general overview of the cerebellar adaptive mechanisms (they are biologically inspired but not biologically plausible). Thus biology is only taken into consideration at a very high level of abstraction. The solutions provided are usually purely speculative and difficult to refute/validate from a cellular/neural network point of view. These solutions aim at obtaining performance in the robotic VOR task itself rather than understanding the biological involvements. The most prominent examples found in this machine learning family are:

- *Learning systems derived from the biologically inspired principle of feedback-error learning* (FEL) [40] combined with non-parametric statistical learning networks [41]. FEL approximately maps the sensory error into motor error. The motor error is subsequently used to train a neural network through supervised learning by means of a recursive least squares algorithm (RLS) based on a Newton-like method. RLS facilitates a very fast convergence and robustness without the need for costly parameter adjustments. This system is able to acquire a high performance visual stabilization reflex in a humanoid robot but the biological plausibility is lacking.

- *Learning systems based on adaptive linear filters as cerebellar controllers*. The Marr-Albus theory commonly assumes the teaching signal (from CFs) as the motor error. This assumption demands complex neural structures that are able to estimate non-observable motor errors from their observable sensory consequences. To that aim, a recurrent control architecture with a controller that decorrelates the sensory error from the motor error is used [42]. These learning systems assume the cerebellum operating like a bank of adaptive linear filters supervised by the CF activity [43].

- *Learning systems based on local weight projection regressions* (LWPR) [44] as cerebellar controllers. LWPR is a non-linear function approximator that operates in high dimensional spaces. This algorithm is able to cope with redundant dimensions and irrelevant inputs. These learning systems use a cerebellar model in which the granular and molecular layers (also including the interneurons [45]) are modeled using this LWPR algorithm [46, 47]. The input to the PCs is the output of the LWPR algorithm. This cerebellar model has been used to create a gaze stabilization system in [48].

B. *The cerebellar-based family*

The cerebellar-based family solves the embodied cognition approach (VOR) by taking the biological restrictions imposed by the cerebellar neural structures as granted. The cerebellar algorithm performance is a consequence of the built-in biologically plausible integrated characteristics, not the main target. The cerebellar algorithms are biologically constrained and they share a family resemblance with the cerebellar anatomy (they aim to be biologically inspired and biologically plausible). The solutions provided give us a closer and clearer view of the cerebellar computation primitives. The main aim here is to draw humanoid-human analogies that may drive basic cerebellar research by proposing working hypotheses that can be either refuted or validated from a cellular or neural network point of view. This family can be sub-divided into two main categories: analog cerebellar models and spiking cerebellar models.

- *Analog cerebellar models.* These models usually present higher abstraction levels than spiking models (assuming rate coding at cell level representation). They are usually easier to implement and more computationally efficient at the expense of being less biologically plausible. These kinds of cerebellar models have been used to recreate an eye blink classic conditioning (EBCC) [49] and a VOR experiment [50]. RT requirements here are easy to cope with due to the simplicity and efficiency of the analog cerebellar model.

- *Spiking cerebellar models.* These models are more akin to biology. They try to mimic the cerebellar neural communication by using spikes (thus, even spatio-temporal spiking representations and STDP mechanisms can be studied). Spikes are propagated within cerebellar sub-circuits that attempt to mimic the cerebellar architecture. Interestingly, the emerging behavior from the dialogue between the neural code and the different cerebellar sub-circuits is intended to cope with the behaviors observed in biology. These spiking models can be designed using the results obtained in experimental neuroscience to increase their biological plausibility. These complex models can then be used conjointly with experimental neuroscience to easily refute/validate new hypothesis that could hardly be studied just by experimental neuroscience due to its inherent technical limitations. Nevertheless, conciliating realistic spiking cerebellar models with behavioral outcomes (i.e. VOR) remains an open issue. Computational models that partly address this problem exist (i.e. modeling and interconnecting certain sub-circuits [51] or certain spiking features [52]). Nevertheless, reconstructing the path from cellular to behavior level remains elusive. ***To the best of our knowledge, the solution proposed in this study is one of the first initiatives that are able to combine this level of neural detail with several neural adaptive mechanisms all working together to operate a humanoid performing a VOR experiment in RT***. In this case, the RT requirements are harder to cope with due to the higher complexity of the spiking cerebellar model.

## V. Conclusion

In this study, *we present one of the first cerebellar embodiment case-of-studies able to effectively reproduce an r-VOR task with a real humanoid robot in RT*. The spiking cerebellar model/controller effectively adapts the reflex for a real iCub robot thanks to the two STDP mechanisms located at the PF-PC and MF-VN synapses. Both STDP mechanisms operate conjointly to shape the cerebellar neural activity that ultimately generates the eye motor commands that compensate for the head movement in the iCub robot.

This case-of-study incorporates two key elements, for the first time in cerebellar embodiment, which are pivotal to establish a coherent communication between the cerebellar controller/model and the front-end body (iCub) in RT; (1) an inner/outer control loop and (2) a RT supervisor. These two elements solve the body-mind dialog technical problem in RT thus ensuring a proper timing between the spiking cerebellar commands generated and their corresponding motor actions/sensory responses.

Acknowledgments

We would like to thank the Institut des Systèmes Intelligents et de Robotique (ISIR), from Sorbonne University, Paris, France and especially Ryan Lober and Jorhabib Eljaik (Sorbonne Université, CNRS UMR 7222, ISIR, F-75005 Paris, France) for configuring and sharing their iCub robot.

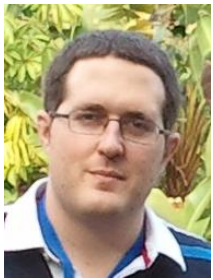

**Francisco Naveros** received two M.Sc. degrees, one in Telecom Engineering and the other in Computer Science and Networks as well as a Ph.D. degree in Computational Neuroscience from the University of Granada (Spain), in 2011, 2012 and 2017 respectively.

Since 2017, he has been a postdoctoral researcher in the Computational Neuroscience and Neurorobotics Lab of the University of Granada. He is the author of 5 articles. His main research interests include biologically processing control schemes, parallel and real-time spiking neural network simulations and light weight robots.

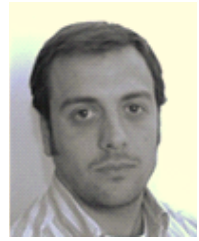

**Niceto R. Luque** was awarded his M.Sc. and Ph.D. degrees in Computer Science and Networks from the University of Granada (Spain) in 2007 and 2013 respectively. He also received a B.Sc in Electronic Engineering and a M.Sc. in Automatics and Industrial Electronics from the University of Córdoba (Spain) in 2003 and 2006, respectively.

From 2015 to 2017, he obtained an IF Marie Curie Post-Doc Fellowship from the EU Commission in Dr. Arleo's lab in Paris. In 2018 he obtained a Juan de la Cierva Incorporation Post-Doc Fellowship from the Spanish Government in the Computational Neuroscience and Neurorobotics Lab ofthe University of Granada. He is the author of more than 20 articles. His main research interests include biologically processing control schemes, light weight robots, spiking neural networks and ageing.

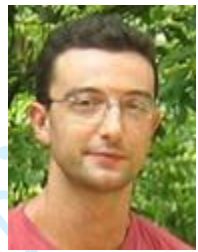

**Eduardo Ros** received his M.Sc. and Ph.D. degrees in Physics and Computational Neuroscience from the University of Granada (Spain) in 1992 and 1997 respectively.

He is currently Full Professor in the Department of Computer Architecture and Technology of the University of Granada. He is the head of the Computational Neuroscience and Neurorobotics Lab of the University of Granada. He is the author of more than 85 articles. His main research interests include bio-inspired processing and neuromorphic engineering, spiking neural networks, hardware processing architectures and computational neuroscience.

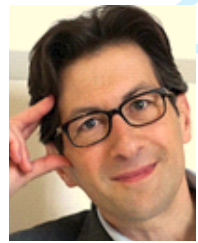

**Angelo Arleo** received his M.Sc. from the University of Mathematical Science of Milan (Italy) in 1996, his Ph.D. in Computational Neuroscience from the EPFL (Switzerland) in 2000, and his Habilitation to Direct Research (HDR) from the University Pierre & Marie Curie, Paris (France) in 2005.

In 2014, he joined the Vision Institute to explore the perceptual and cognitive consequences of visual aging in humans. Overall, his research focused on the neural processes mediating multisensory integration, contextual learning, and spatial cognitive functions. He always combined experimental and computational neuroscience to cross-link multiple organization levels, and he has a track record in life science, computational neurobiology, and neuro-engineering.

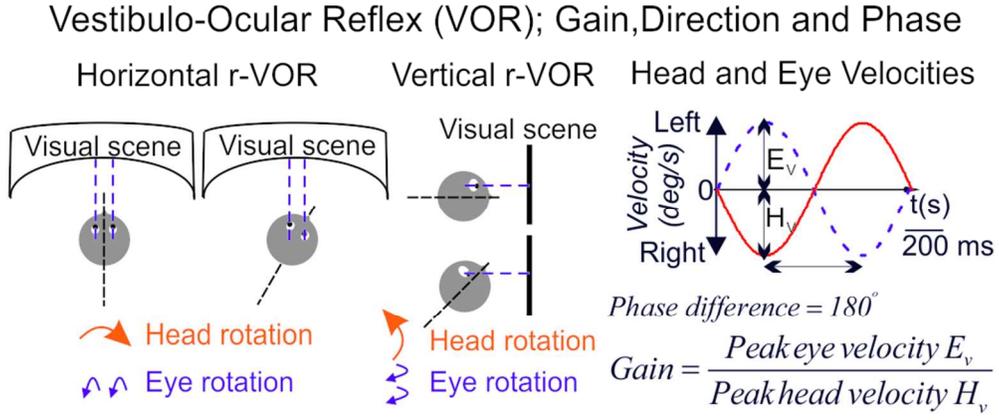

Fig. 1. Vestibulo-Ocular Reflex (VOR) experiment. VOR stabilizes the images on the fovea during horizontal and vertical head rotation tests by producing opposite eye movements that compensate the movement.

88x37mm (300 x 300 DPI)

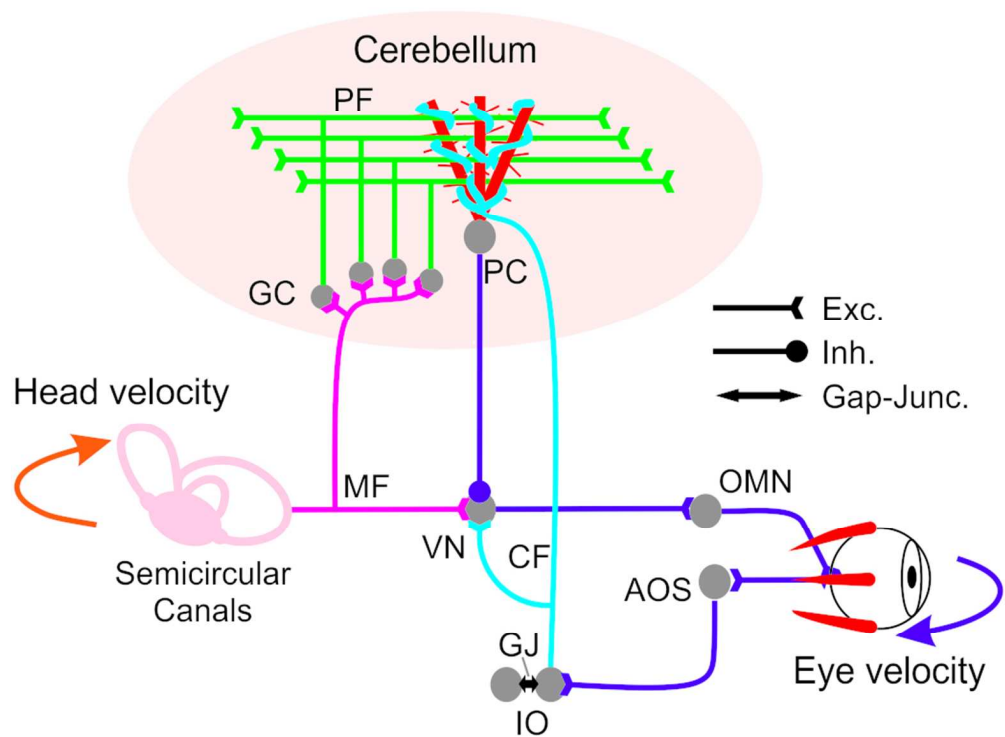

Fig. 2. Vestibular and cerebellar scheme. Connections from semicircular canals in vestibular organ to oculomotor nucleus (OMN) via the flocculus in the cerebellum and the vestibular nuclei (VN), forming the three-neuron reflex arc (MF: mossy fibers, GC: granular cells, PF: parallel fibers, PC: Purkinje cells, CF: climbing fibers, IO: inferior olive, GJ: gap-junction, AOS: accessory optic system).

87x64mm (300 x 300 DPI)

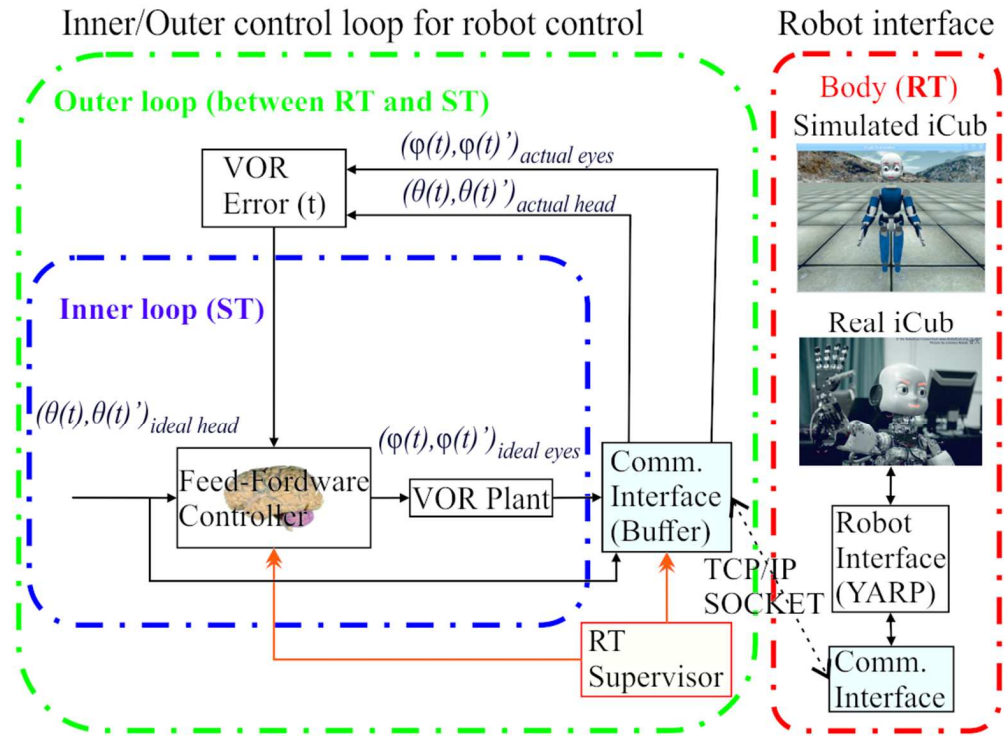

Fig. 3. Cerebellar inner/outer control loop and robot interface for r-VOR tasks. The vestibular and teaching (error) neural signals arrive to the cerebellar model through the MFs and IOs respectively (after an analog-to-spike conversion). The cerebellar model processes this input activity and generates the corresponding output response through the VN. This VN output spike activity is transformed into an analog signal that feeds the VOR plant. The VOR plant output, representing the eye velocity commands, is stored in the outer loop buffer. The robot interface module concurrently connects with the outer loop communication interface, receiving the eye velocity commands from the inner loop and sending the sensory information from the iCub sensors (head and eyes trajectories). Then, the outer loop compares both trajectories, generating the teaching signal (IO activity) in the VOR error module. The RT supervisor manages the simulation speed and the synchronization between both elements using a temporal buffer.

90x66mm (300 x 300 DPI)

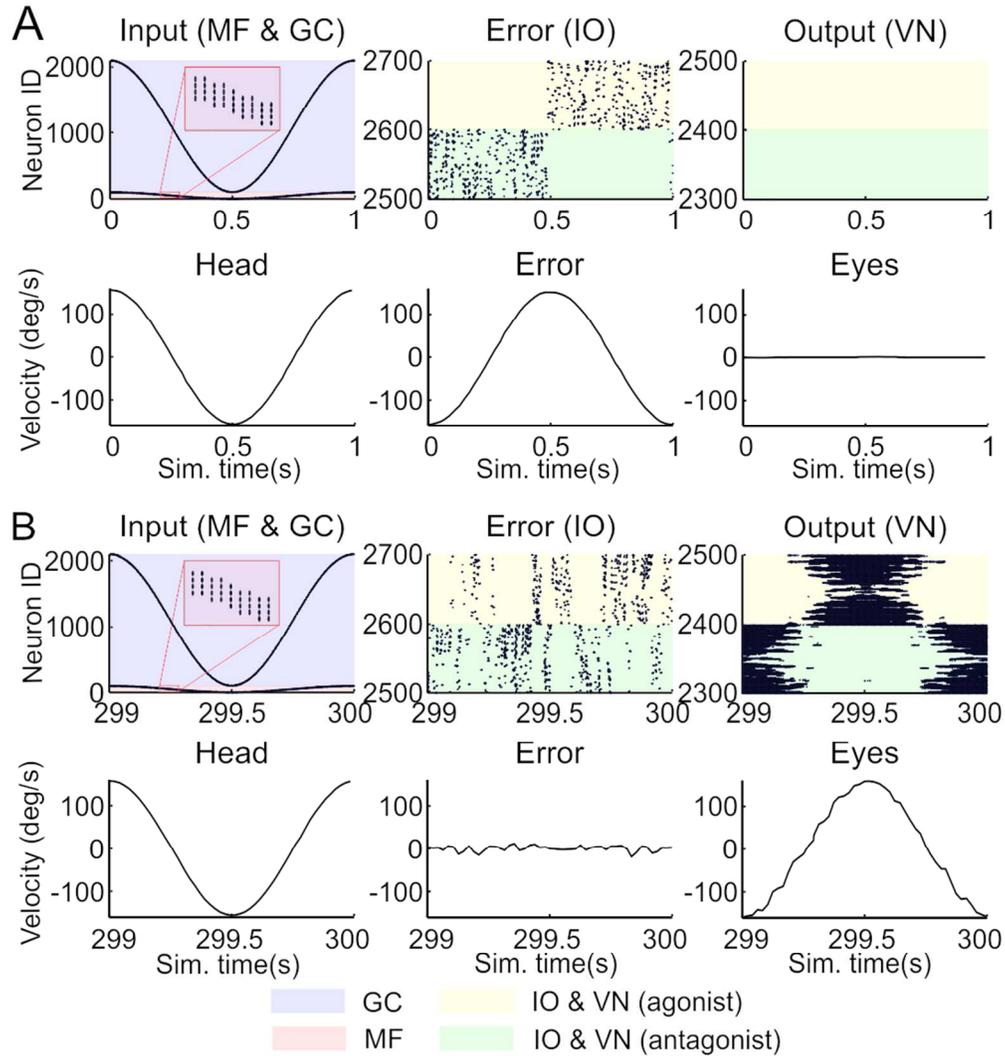

Fig. 4. Cerebellar input/output signals during horizontal r-VOR task (150 deg/s). A) Initial learning stage. B) Final learning stage. The first row of each panel depicts the spiking cerebellar input/output signals, whereas the second row shows their analog translation. The left column depicts the cerebellar input (head velocity), the central column the error input (mismatch between head and eyes velocities) and the right column the cerebellar output (eyes velocity with respect to head movement). The right column analog signal actually represents the cerebellar output emerging from the VOR plant.

91x97mm (300 x 300 DPI)

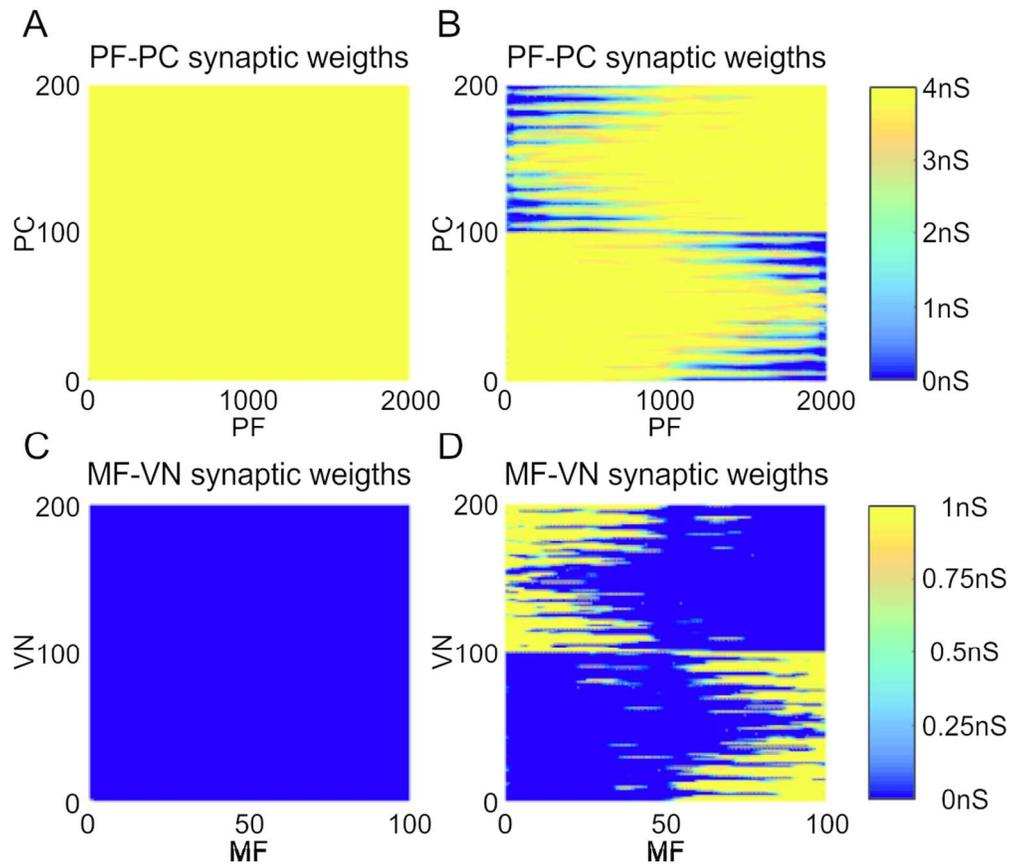

Fig. 5. Synaptic weight distribution during a horizontal r-VOR task (150 deg/s). The first row depicts synaptic weight distribution at the PF-PC innervations. The second row depicts the synaptic weight distribution at the MF-VN innervations. For each row, the left hand column represents the synaptic weight distributions at the initial learning stage, whilst the right hand column represents the synaptic weight distributions at the final learning stage.

105x90mm (300 x 300 DPI)

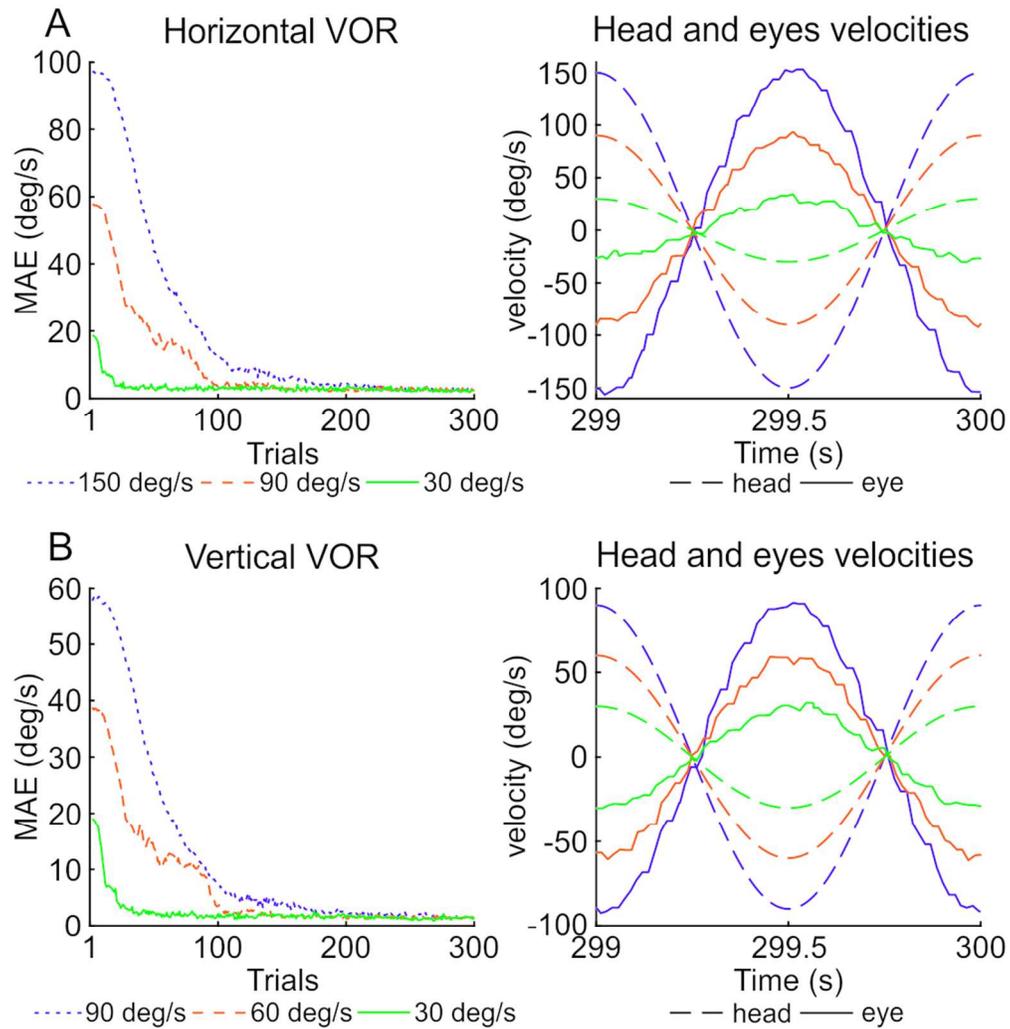

Fig. 6. Cerebellar output response to horizontal and vertical r-VOR tasks. A) Horizontal r-VOR tasks (150, 90 and 30 deg/s). B) Vertical r-VOR tasks (90, 60 and 30 deg/s). The left hand column represents the mean absolute error (MAE) evolution between the velocities of the head and eyes during the 300 trials (1 second per trial) of each experiment. The right hand column represents curves of the head and eyes during the last trial of each task.

93x96mm (300 x 300 DPI)

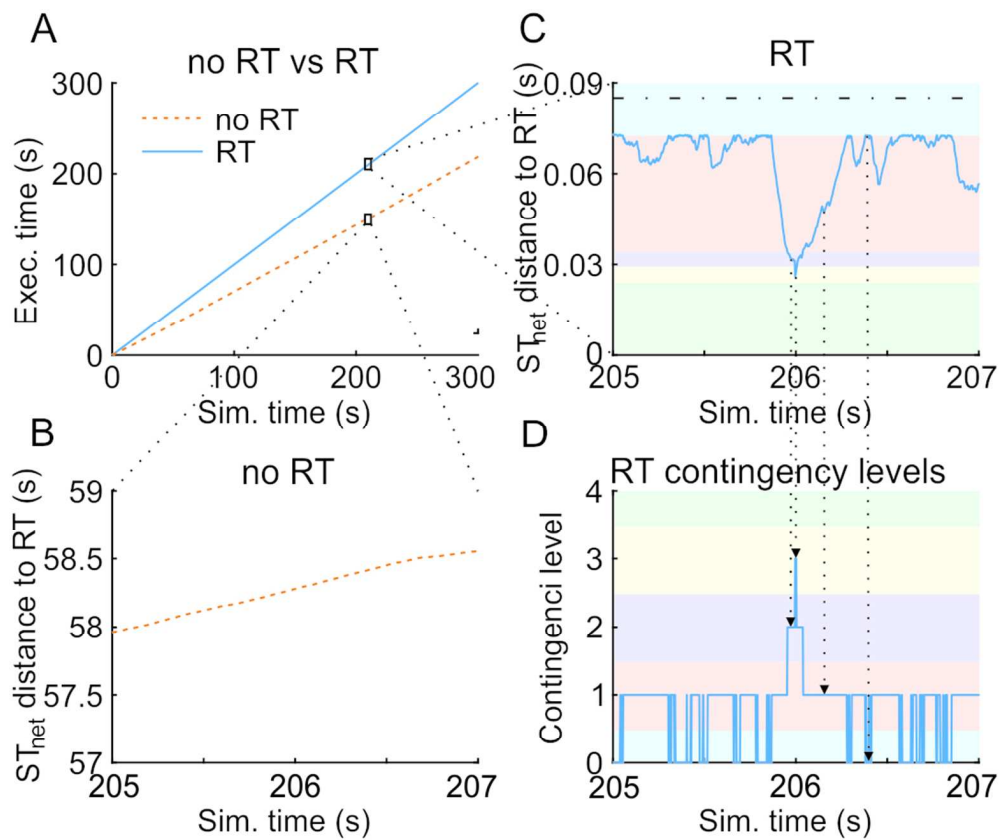

Fig. 7. RT supervisor impact on the neural simulation time speed in a horizontal r-VOR task (comparison when enabling and disabling the RT supervisor). A) Cerebellar simulation time speed (ratio between simulation and execution time). B) and C) Cerebellar simulation time (STnet) distance to RT. This distance must not surpass 85 ms to meet the RT bounds required for a coherent sensory-motor propagation. D) RT supervisor contingency level depending on the distance between STnet and RT.

91x76mm (300 x 300 DPI)

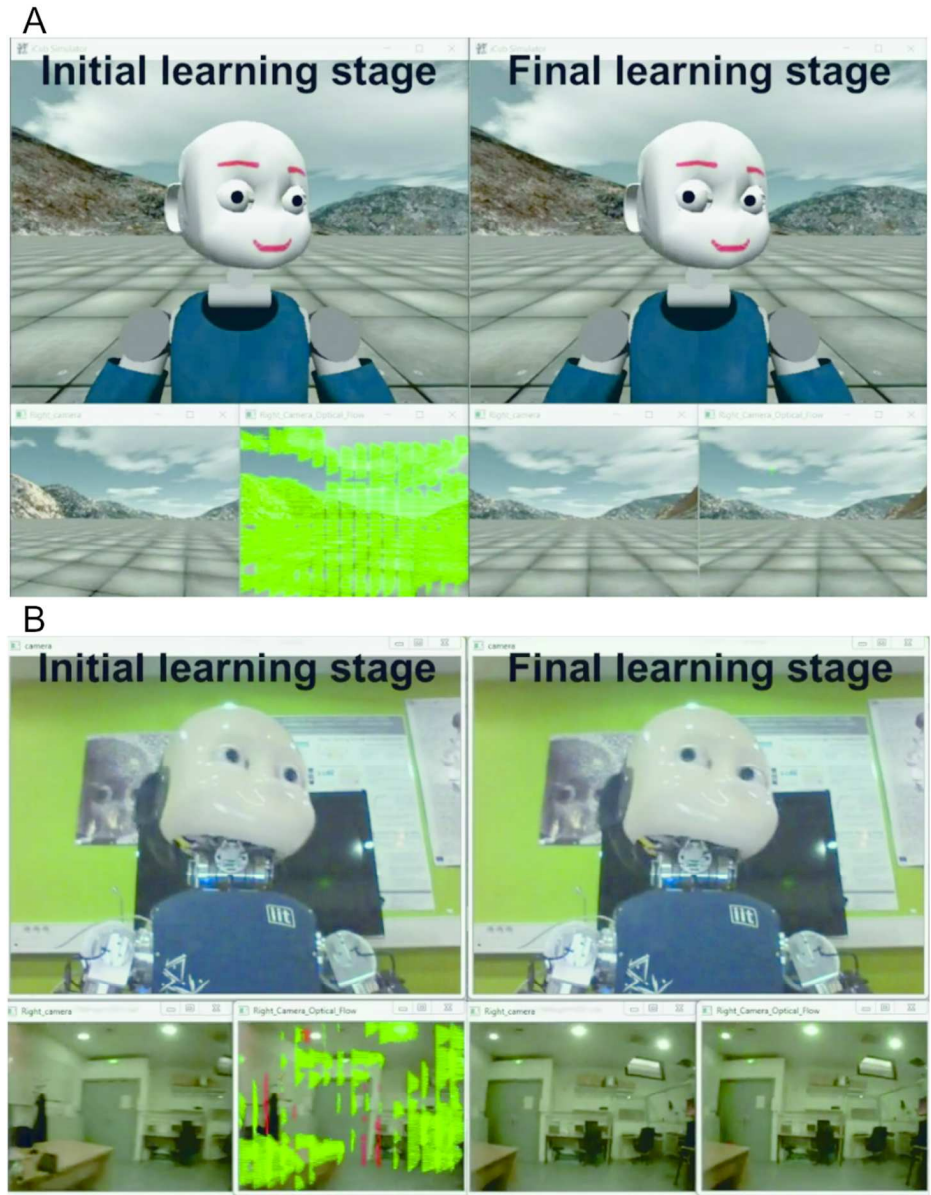

Fig. 8. Snapshot of the movies filming the r-VOR in the simulated (A) and real (B) iCub robot. The left hand windows represent the initial learning stage whereas the right hand windows represent the final learning stage. The upper windows in A and B show the head and eyes movements whereas the lower windows show the images filmed by the eye cameras. The optical flow is computed over the camera images (superimposed in green arrows) indicating quantitatively the level of stabilization of the filmed image on the "retina".

93x120mm (300 x 300 DPI)